\documentstyle[12pt]{article}

\newcommand{\sect}[1]{\setcounter{equation}{0}\section{#1}}

\def\be{\begin{equation}}
\def\ee{\end{equation}}
\def\ba{\begin{eqnarray}\samepage}
\def\ea{\end{eqnarray}}

\font\twelvemsa=msam10 scaled 1200

\font\sevenmsa=msam7
\font\fivemsa=msam5
\newfam\msafam
\textfont\msafam=\twelvemsa
\scriptfont\msafam=\sevenmsa
\scriptscriptfont\msafam=\fivemsa
\def\msa{\ifcase\msafam 0\or1\or2\or3\or4\or5\or6\or7\or8\or9\or A\or B\or
C\or D\or E\or F\fi}

\font\twelvemsb=msbm10 scaled 1200
\font\elevenmsb=msbm10 scaled 1100

\font\sevenmsb=msbm7
\font\fivemsb=msbm5
\newfam\msbfam
\textfont\msbfam=\twelvemsb
\scriptfont\msbfam=\sevenmsb
\scriptscriptfont\msbfam=\fivemsb
\def\msb{\ifcase\msbfam 0\or1\or2\or3\or4\or5\or6\or7\or8\or9\or A\or B\or
C\or D\or E\or F\fi}

\font\twelveeuf=eufm10 scaled 1200

\font\seveneuf=eufm7
\font\fiveeuf=eufm5
\newfam\euffam
\textfont\euffam=\twelveeuf
\scriptfont\euffam=\seveneuf
\scriptscriptfont\euffam=\fiveeuf
\def\euf{\ifcase\euffam 0\or1\or2\or3\or4\or5\or6\or7\or8\or9\or A\or B\or
C\or D\or E\or F\fi}

\def\goth#1{\fam\euffam#1}
\def\Bbb#1{\fam\msbfam#1}
\def\fraction#1#2{{\textstyle\frac{#1}{#2}}}
\def\half{\textstyle\frac{1}{2}}

\mathchardef\gapprox"3\msa26
\mathchardef\lapprox"3\msa2E

\begin{document}

\title{SL(2,R) INVARIANCE OF NON-LINEAR ELECTRODYNAMICS COUPLED TO AN
AXION AND A DILATON}

\author{G W GIBBONS\\ \&\\D A RASHEED\thanks{Supported by EPSRC grant
no.~9400616X.} \\ \\D.A.M.T.P.\\University of
Cambridge\\Silver Street\\Cambridge CB3 9EW\\U.K.}

\maketitle

\begin{abstract}
\noindent
The most general Lagrangian for non-linear electrodynamics coupled to
an axion $a$ and a dilaton $\phi$ with $SL(2,\mbox{\elevenmsb R})$
invariant equations of motion is
$$
-\half\left(\nabla\phi\right)^2 - \half e^{2\phi}\left(\nabla
a\right)^2 + \fraction{1}{4}aF_{\mu\nu}\star F^{\mu\nu} + L_{\rm
inv}(g_{\mu\nu},e^{-\frac{1}{2}\phi}F_{\rho\sigma})
$$
where
$L_{\rm inv}(g_{\mu\nu},F_{\rho\sigma})$ is a Lagrangian
whose equations of motion are invariant under electric-magnetic
duality rotations. In particular there is a unique generalization of
Born-Infeld theory admitting $SL(2,\mbox{\elevenmsb R})$ invariant
equations of motion.
\end{abstract}

\renewcommand{\thepage}{ }
\pagebreak

\renewcommand{\thepage}{\arabic{page}}
\setcounter{page}{1}

\sect{Introduction}

In a recent paper \cite{GibRas} we found the condition on the
Lagrangian function $L_{\rm inv}(g_{\mu\nu},F_{\rho\sigma})$ for a
non-linear electrodynamic theory coupled to gravity that the
equations of motion, including the Einstein equations, are invariant
under the action of an $SO(2)$ group of generalized
electric-magnetic duality rotations. One such Lagrangian is the
Born-Infeld Lagrangian \cite{BI}
\be
L_{\rm BI} = 1 - \sqrt{1+\fraction{1}{2}F^2-\fraction{1}{16}(F\star
F)^2}.
\ee
Note that $L_{\rm inv}(g_{\mu\nu},F_{\rho\sigma})$ itself
is {\em not} invariant under duality rotations.

In this letter we shall extend these results by including a coupling
to a scalar dilaton field $\phi$ and a pseudo-scalar axion field
$a$. They contribute to the action the following kinetic terms
\be
L_{\rm ax-dil} = -\half\left(\nabla\phi\right)^2 -
\half e^{2\phi}\left(\nabla a\right)^2
\ee
which are $SL(2,{\Bbb R})$ invariant. We shall show that this
$SL(2,{\Bbb R})$ invariance may be extended to the equations of motion
(but not the action) if and only if the action takes the form
\be
\int d^4x\sqrt{g} \left\{R - 2\Lambda -
\half\left(\nabla\phi\right)^2 - \half e^{2\phi}\left(\nabla
a\right)^2 + \fraction{1}{4}aF\star F +
L_{\rm inv}(g_{\mu\nu},e^{-\frac{1}{2}\phi}F_{\rho\sigma})
\right\}
\ee
where $L_{\rm inv}\left(g_{\mu\nu},F_{\rho\sigma}\right)$ is a
Lagrangian with $SO(2)$ invariant equations of motion. In particular
there is just one generalization of the Born-Infeld Lagrangian
admitting $SL(2,{\Bbb R})$ invariant equations of motion. This
$SL(2,{\Bbb R})$ invariant generalization of the Born-Infeld
Lagrangian does {\em not} coincide with that discussed in
\cite{GibRas} in connection with string theory. The relation of our
new results to string theory is currently under investigation.

\sect{SL(2,R) Duality at Lowest Order}

In 4 dimensions, the bosonic sector of N=4 supergravity and string
theory compactified on a torus, at lowest order, may be described by
the following Lagrangian \cite{CreSchFer}
\be
L = R - \half\left(\nabla\phi\right)^2 - \half e^{2\phi}\left(\nabla
a\right)^2 + \fraction{1}{4}aF_{\mu\nu}\star F^{\mu\nu} -
\fraction{1}{4}e^{-\phi}F_{\mu\nu}F^{\mu\nu}
\label{LowEn}
\ee
where for simplicity we consider only a single $U(1)$ gauge
field\footnote{In this paper we use units in which
$\mu_0=\varepsilon_0=\hbar=c=16\pi G=1$.}. The resulting theory admits
an $SL(2,{\Bbb R})$ electric-magnetic duality which mixes the
electromagnetic field equations with the Bianchi identities and also
transforms the axion and dilaton.

In a local orthonormal frame, the electric intensity $\bf E$ and
magnetic induction $\bf B$ may be defined by $E_i=F_{i0}$ and $B_i={1
\over 2}\epsilon_{ijk}F_{jk}$. The Bianchi identities $dF=0$ or
$\partial_{[\alpha} F_{\beta\gamma]}=0$ are then equivalent to
\ba
{\bf\nabla}\cdot{\bf B} & = & 0 \nonumber \\
 & & \\
{\bf\nabla}\times{\bf E} & = & - {\partial{\bf B}\over\partial t}.
\nonumber
\ea
Defining $G^{\mu\nu}$ by\footnote{There is
some ambiguity in the definition of this partial derivative depending on
whether or not one takes into account the antisymmetry of $F_{\mu\nu}$.
Here we treat $F_{\mu\nu}$ and $F_{\nu\mu}$ as independent variables,
hence the factor of 2.}
\be
G^{\mu\nu} = -2 {\partial L\over\partial F_{\mu\nu}},
\label{Constit}
\ee
the field equations are $d\star G=0$ or $\partial_{[\alpha}\star
G_{\beta\gamma]}=0$,
which are equivalent to
\ba
{\bf\nabla}\cdot{\bf D} & = & 0 \nonumber \\
 & & \\
{\bf\nabla}\times{\bf H} & = & + {\partial{\bf D}\over\partial t},
\nonumber
\ea
where the electric induction $\bf D$ and magnetic intensity $\bf H$
are defined by $D_i=G_{i0}$ and $H_i={1\over
2}\epsilon_{ijk}G_{jk}$. For the Lagrangian (\ref{LowEn}), $\bf D$ and
$\bf H$ are given by
\ba
{\bf D} & = & + {\partial L\over\partial{\bf E}} = e^{-\phi}{\bf E} +
a{\bf B} \nonumber \\
 & & \label{LowEnDH}\\
{\bf H} & = & - {\partial L\over\partial{\bf B}} = e^{-\phi}{\bf B} -
a{\bf E} \nonumber
\ea
These are the constitutive relations and may be rewritten as
\be
\left({{\bf E}\atop{\bf H}}\right) =
\underbrace{
\left(
\begin{array}{cc}
e^\phi & -ae^\phi \\
-ae^\phi & e^{-\phi}+a^2e^\phi
\end{array}
\right)}
_{\textstyle\cal M}
\left({{\bf D}\atop{\bf B}}\right).
\label{SL2RConstit}
\ee
It is convenient to define a complex scalar field
$\lambda$ by
\be
\lambda = a + ie^{-\phi}
\ee
and a complex 2-component vector $\psi$ by
\be
\psi = \left({1\atop -\lambda}\right).
\ee
Then the matrix ${\cal M}$ may be written as
\be
{\cal M} = {\psi\psi^\dagger + c.c.\over\sqrt{{\rm
det}\left(\psi\psi^\dagger + c.c.\right)}}.
\ee
Chosing the first component of $\psi$ to be $1$ fixes the representation
of ${\cal M}$. The $SL(2,{\Bbb R})$ duality transformation may then be
constructed so that it automatically leaves the constitutive relations
invariant~:
\ba
\psi \rightarrow \psi^\prime \propto \left(S^T\right)^{-1}\psi \quad &
\Rightarrow & \quad {\cal M} \rightarrow \left(S^T\right)^{-1}{\cal
M}S^{-1} \nonumber \\
 & & \\
\left({{\bf D}\atop{\bf B}}\right) \rightarrow S\left({{\bf D}\atop{\bf
B}}\right) \quad & , & \quad \left({{\bf E}\atop{\bf H}}\right)
\rightarrow\left(S^T\right)^{-1}\left({{\bf E}\atop{\bf H}}\right),
\nonumber
\ea
where $S\in SL(2,{\Bbb R})$. If
\be
S = \left(
\begin{array}{cc}
p & q \\
r & s
\end{array}
\right)
\qquad\mbox{where  }ps-qr=1,
\ee
then the induced transformations of the axion and dilaton fields are
given by a Mobius transformation of $\lambda$~:
\be
\lambda \rightarrow {p\lambda+q\over r\lambda+s}.
\label{axdiltransf}
\ee
It is easy to check that the axion and dilaton equations of motion are
invariant under these transformations and also so is the energy momentum
tensor, so it is consistent to assume that the metric is unchanged
under the action of this duality.

In the covariant notation, the transformations of $F$ and $G$ are given by
\be
\left\{
\begin{array}{rcl}
F_{\mu\nu} & \rightarrow & s F_{\mu\nu} + r\star G_{\mu\nu} \\
 & & \\
G_{\mu\nu} & \rightarrow & p G_{\mu\nu} - q\star F_{\mu \nu}.
\end{array}
\right.
\ee
Defining the complex 2-forms ${\goth F}=F+i\star F$ and ${\goth
G}=\star G-iG$, the duality may be written more compactly as~:
\be
\left({{\goth G}\atop{\goth F}}\right)
\rightarrow
\left(
\begin{array}{cc}
p & q \\
r & s
\end{array}
\right)
\left({{\goth G}\atop{\goth F}}\right)
\quad,\quad
\lambda \rightarrow {p\lambda+q\over r\lambda+s}
\quad,\quad
g_{\mu\nu}\rightarrow g_{\mu\nu}.
\label{Duality}
\ee

\sect{SL(2,R) Duality in Non-Linear Electrodynamics}

Since in both string theory and in supergravity theories, higher order
terms in the electromagnetic field arise, causing the
electrodynamic equations of motion to become non-linear, it is natural
to ask under what circumstances the $SL(2,{\Bbb R})$ duality above
continues to hold. It has been shown \cite{GibRas} that, in the case
of pure non-linear electrodynamics with no axion or dilaton, the
equations of motion will admit an $SO(2)$ duality provided the
Lagrangian satisfies a simple differential constraint~:
$G_{\mu\nu}\star G^{\mu\nu}=F_{\mu\nu}\star F^{\mu\nu}$, or
equivalently ${\bf E}\cdot{\bf B}={\bf D}\cdot{\bf H}$. Moreover there
are, roughly speaking, as many Lagrangians satisfying this constraint
as there are functions of a single real variable. Amongst this class
of Lagrangians is the Born-Infeld Lagrangian~:
\be
\sqrt{g}L = \sqrt{g} - \sqrt{{\rm det}(g_{\mu\nu}+F_{\mu\nu})},
\ee
which, in 4 dimensions gives
\be
L = 1 - \sqrt{1+\fraction{1}{2}F^2-\fraction{1}{16}(F\star F)^2}.
\ee

We will consider here a 4 dimensional Lagrangian $L(g,F,a,\phi)$ which
has the same axion an dilaton kinetic terms as (\ref{LowEn}) but we
will allow an arbitrary dependence on $a$, $\phi$ and $F_{\mu\nu}$. We
will not consider the higher order derivative terms in $F$ which also
occur in string theory. There is some evidence that the singularities
present in solutions of the Einstein-Maxwell theory are absent in
string theory due to higher order corrections and consequently the
higher order derivative terms in $F$ may be neglected \cite{Tse}.

Infinitesimally, the $SL(2,{\Bbb R})$ transformation above may be
described by the matrix
\be
S = \left(
\begin{array}{cc}
p & q \\
r & s
\end{array}
\right) = \left(
\begin{array}{cc}
1+\alpha & \beta \\
\gamma & 1-\alpha
\end{array}
\right).
\ee
The fields then transform according to
\be
\left\{
\begin{array}{rl}
\delta a & = 2\alpha a + \beta -\gamma(a^2-e^{-2\phi}) \\
\delta\phi & = 2(a\gamma - \alpha) \\
\delta F_{\mu\nu} & = \gamma\star G_{\mu\nu} - \alpha F_{\mu\nu} \\
\delta G_{\mu\nu} & = \alpha G_{\mu\nu} - \beta\star F_{\mu\nu} \\
\delta g_{\mu\nu} & = 0.
\end{array}
\right.
\label{Transf}
\ee

\subsection{Invariance of constitutive relations}

Invariance of the constitutive relation under these transformations
requires that
\be
\delta G^{\mu\nu} = -2{\partial\over\partial
F_{\rho\sigma}}\left({\partial L\over\partial F_{\mu\nu}}\right)\delta
F_{\rho\sigma} -2{\partial\over\partial a}\left({\partial
L\over\partial F_{\mu\nu}}\right)\delta a
-2{\partial\over\partial\phi}\left({\partial L\over\partial
F_{\mu\nu}}\right)\delta\phi.
\label{Star}
\ee
Comparing the coefficients of $\alpha$, $\beta$ and $\gamma$ in this
equation gives 3 differential constraints on the Lagrangian. Firstly,
the $\beta$ equation reads
\be
{\partial^2L\over\partial a\partial F_{\mu\nu}} = {1\over 2}\star
F^{\mu\nu}.
\ee
Integrating this implies that $L$ must be of the form
\be
L = R - \half\left(\nabla\phi\right)^2 - \half e^{2\phi}\left(\nabla
a\right)^2 + \fraction{1}{4}aF_{\mu\nu}\star F^{\mu\nu} +
\widetilde{L}(F,\phi) + f(a,\phi).
\ee
The coefficients of $\alpha$ in (\ref{Star}) then give the following
constraint on $\widetilde{L}$~:
\be
{\partial\over\partial
F_{\rho\sigma}}\left(F_{\rho\sigma}{\partial\widetilde{L}\over\partial
F_{\mu\nu}}\right) + 2{\partial^2\widetilde{L}\over\partial\phi\partial
F_{\mu\nu}} = 0.
\ee
Defining a new 2-form field
$\bar{F}_{\mu\nu}=e^{-\frac{1}{2}\phi}F_{\mu\nu}$ and
changing variables to $\bar{F}$ and $\phi$ in $\widetilde{L}$, this
last constraint reads
\be
{\partial^2\widetilde{L}\over\partial\phi\partial\bar{F}_{\mu\nu}} = 0
\ee
which implies that $\widetilde{L}=\widetilde{L}(e^{-\frac{1}{2}\phi}F)$
plus an arbitrary function of $\phi$ which we are free to absorb into
$f(a,\phi)$. The coefficients of $\gamma$ in (\ref{Star}) then give
another constraint on $\widetilde{L}$~:
\be
{\partial\over\partial\bar{F}_{\mu\nu}}\left(
{\partial\widetilde{L}\over\partial\bar{F}_{\rho\sigma}}
{\partial\widetilde{L}\over\partial\bar{F}_{\lambda\tau}}\right)
\eta_{\rho\sigma\lambda\tau} = {1\over
2}\eta^{\mu\nu\rho\sigma}\bar{F}_{\rho\sigma},
\ee
where $\eta_{\mu\nu\rho\sigma}$ is the completely antisymmetric
{\em tensor} of the 4 dimensional spacetime. It is natural at this
stage to define
\be
\bar{G}^{\mu\nu} =
-2{\partial\widetilde{L}\over\partial\bar{F}_{\mu\nu}} =
e^{\frac{1}{2}\phi}(G^{\mu\nu}+a\star F^{\mu\nu}).
\label{BarG}
\ee
Then the last constraint can be integrated to give
\be
\bar{G}_{\mu\nu}\star\bar{G}^{\mu\nu} =
\bar{F}_{\mu\nu}\star\bar{F}^{\mu\nu} + 4C
\label{Cond1}
\ee
where C is an arbitrary constant.

\subsection{Invariance of axion and dilaton equations}

The dilaton and axion equations of motion are respectively
\be
-\nabla^2\phi = -e^{2\phi}(\nabla a)^2 +
\fraction{1}{4}F_{\mu\nu}(G^{\mu\nu}+a\star F^{\mu\nu}) + {\partial
f\over\partial\phi}
\label{Dil}
\ee
and
\be
-\nabla^2 a = 2(\nabla a)(\nabla\phi) +
\fraction{1}{4}e^{-2\phi}F_{\mu\nu}\star F^{\mu\nu} + {\partial
f\over\partial a}.
\label{Ax}
\ee
These equations are also required to be invariant under the
$SL(2,{\Bbb R})$ transformations and (\ref{Transf}) implies that they
must transform into one another according to
\ba
\delta(\mbox{Eq. (\ref{Dil})}) & = & 2\gamma\,\mbox{Eq. (\ref{Ax})}
\nonumber \\
 & & \\
\delta(\mbox{Eq. (\ref{Ax})}) & = &
2(\alpha-a\gamma)\,\mbox{Eq. (\ref{Ax})} - 2\gamma
e^{-2\phi}\,\mbox{Eq. (\ref{Dil})} \nonumber
\ea
Comparing terms involving $F$ and $G$ and using equation (\ref{Cond1})
implies that the constant $C$ must vanish. Comparing the terms
involving $f(a,\phi)$ implies that $f$ is at most a constant. The
remaining terms then balance. Combining the results so far, we have
narrowed down the choice of possible Lagrangians to those of the form
\be
L = R - 2\Lambda - \half\left(\nabla\phi\right)^2
- \half e^{2\phi}\left(\nabla a\right)^2 +
\fraction{1}{4}aF_{\mu\nu}\star F^{\mu\nu} +
\widetilde{L}(e^{-\frac{1}{2}\phi}F)
\label{Lag}
\ee
where $\widetilde{L}(\bar{F})$ is required to satisfy
\be
\bar{G}_{\mu\nu}\star\bar{G}^{\mu\nu} =
\bar{F}_{\mu\nu}\star\bar{F}^{\mu\nu}
\label{Cond2}
\ee
or equivalently
\be
(G_{\mu\nu}+a\star F_{\mu\nu})(\star G^{\mu\nu}-aF^{\mu\nu}) =
e^{-2\phi}F_{\mu\nu}\star F^{\mu\nu}.
\label{Cond3}
\ee
In terms of $\bf E$, $\bf B$, $\bf D$ and $\bf H$ this condition reads
\be
({\bf D}-a{\bf B})\cdot({\bf H}+a{\bf E}) = e^{-2\phi}{\bf E}\cdot{\bf
B}
\ee
which is clearly satisfied by the $\bf D$ and $\bf H$ fields defined
in (\ref{LowEnDH}) for the Lagrangian (\ref{LowEn}).

\subsection{Invariance of energy-momentum tensor}

The final point to be checked is that the energy-momentum tensor is
invariant under this action of $SL(2,{\Bbb R})$, otherwise it would
not be consistent to assume that the metric is invariant, which we
have already implicitly done in a number of the steps above. The
energy-momentum tensor is most conveniently defined as
\be
T^{\mu\nu} = g^{\mu\nu}L - {\partial L\over\partial(\partial_\mu
A_\lambda)}(\partial^\nu A_\lambda) - {\partial
L\over\partial(\partial_\mu\phi)}(\partial^\nu\phi) - {\partial
L\over\partial(\partial_\mu a)}(\partial^\nu a)
\ee
which gives\footnote{Note that this is indeed symmetric~: $L$ depends
on $F$ only via the two invariants $F_{\mu\nu}F^{\mu\nu}$ and
$F_{\mu\nu}\star F^{\mu\nu}$. Therefore $G_{\mu\nu}$ will contain only
terms proportional to $F_{\mu\nu}$ and $\star F_{\mu\nu}$, so
${G_\mu}^\lambda F_{\nu\lambda}$ will contain only terms proportional
to ${F_\mu}^\lambda F_{\nu\lambda}$ and ${F_\mu}^\lambda\star
F_{\nu\lambda}={1\over 4}g_{\mu\nu}F^2$ which are both symmetric in
the indices $\mu$, $\nu$.}
\be
T_{\mu\nu} = g_{\mu\nu}L + {G_\mu}^\lambda F_{\nu\lambda} +
(\partial_\mu\phi)(\partial_\nu\phi) + e^{2\phi}(\partial_\mu
a)(\partial_\nu a).
\ee
All the terms involving derivatives of the axion and dilaton are
invariant, since they are the same terms as those that come from the
Lagrangian (\ref{LowEn}). The Lagrangian is not invariant but
transforms according to
\be
\delta L = \fraction{1}{4}aF_{\mu\nu}\star\delta F^{\mu\nu} +
\fraction{1}{4}a\delta F_{\mu\nu}\star F^{\mu\nu} +
\fraction{1}{4}\delta aF_{\mu\nu}\star F^{\mu\nu} +
{\partial\widetilde{L}\over\partial\bar{F}_{\mu\nu}}
e^{-\frac{1}{2}\phi}\left(\delta
F_{\mu\nu} - \half\delta\phi F_{\mu\nu}\right).
\ee
Using (\ref{Transf}), (\ref{BarG}) and (\ref{Cond3}) this gives
\be
\delta L = - \fraction{1}{2}aF_{\mu\nu}G^{\mu\nu} +
\fraction{1}{4}(\beta-\gamma a^2-\gamma e^{-2\phi})F_{\mu\nu}\star
F^{\mu\nu}.
\ee
So, using (\ref{Transf}), the transformation of the energy-momentum
tensor is
\be
\delta T_{\mu\nu} = \gamma{G_\mu}^\lambda\star G_{\nu\lambda} -
\beta\star{F_\mu}^\lambda F_{\nu\lambda} + g_{\mu\nu}\delta L.
\ee
Using (\ref{Cond3}) and the fact that in 4 dimensions, the components
of any 2-form satisfy ${F_\mu}^\lambda\star F_{\nu\lambda}={1\over
4}g_{\mu\nu}F_{\rho\sigma}\star F^{\rho\sigma}$, this variation of
$T_{\mu\nu}$ vanishes as required.

\sect{Conclusions}

We have shown that Lagrangians of the form (\ref{LowEn}) but with
higher order $F$-terms may retain $SL(2,{\Bbb R})$ invariance provided
they are of the form (\ref{Lag}), (\ref{Cond2}). The condition
(\ref{Cond2}) may be recognized as the same condition (in rescaled
variables) that a theory of pure electrodynamics has to satisfy in
order that it admit an $SO(2)$ duality. Thus for every theory of
non-linear electrodynamics described by a Lagrangian $L_{\rm
inv}(g,F)$ which admits an $SO(2)$ duality, we may construct a new
theory with Lagrangian
\be
- \half\left(\nabla\phi\right)^2 - \half e^{2\phi}\left(\nabla
a\right)^2+\fraction{1}{4}aF_{\mu\nu}\star
F^{\mu\nu}+L_{\rm inv}(g,e^{-\frac{1}{2}\phi}F)+\mbox{const.}
\ee
and the new theory will admit an $SL(2,{\Bbb R})$ duality. Thus, as in
\cite{GibRas}, there will be as many such Lagrangians as there are
functions of a single real variable. One such example is the
generalization of the Born-Infeld Lagrangian to include axion and
dilaton fields~:
\be
\begin{array}{c}
L = R - 2\Lambda - \half\left(\nabla\phi\right)^2 -
\half e^{2\phi}\left(\nabla a\right)^2 +
\fraction{1}{4}aF_{\mu\nu}\star F^{\mu\nu} \\
 \\
+ 1 -
\sqrt{1+\fraction{1}{2}e^{-\phi}F^2-\fraction{1}{16}e^{-2\phi}(F\star
F)^2}
\end{array}
\ee
and this will be the only generalization of the Born-Infeld theory
with the $SL(2,{\Bbb R})$ duality (\ref{Duality}).

\end{document}